\documentclass[useAMS,usenatbib,usegraphicx]{mn2e}

\def\kmsec{\mbox{km~s$^{\rm -1}$}}
\def\logg{\mbox{log~{\it g}}}
\def\msun{\mbox{M$_{\odot}$}}
\def\teff{\mbox{$T_{\rm eff}$}}
\def\vt{\mbox{$v_{\rm t}$}}
\def\logrw{\mbox{$\log$(EW/$\lambda$)}}
\def\rpro{\mbox{$r$-process}}
\def\spro{\mbox{$s$-process}}

\def\loggf{$\log gf$}

\def\seg{\mbox{SDSS~J021933.13$+$200830.2}}
\def\segalt{\mbox{SDSS~J021933.13$+$200830.2}}
\def\seggal{\mbox{Segue~2}}
\def\umigal{\mbox{Ursa Minor}}
\def\boogal{\mbox{Bo\"{o}tes~I}}
\def\comgal{\mbox{Coma Berenices}}
\def\hergal{\mbox{Hercules}}
\def\leogal{\mbox{Leo~IV}}
\def\segigal{\mbox{Segue~1}}
\def\wilgal{\mbox{Willman~1}}
\def\umagal{\mbox{Ursa Major~II}}
\def\sgrgal{\mbox{Sagittarius}}
\def\dragal{\mbox{Draco}}

\def\cargal{\mbox{Carina}}
\def\hd{\mbox{HD~122563}}

\def\cs{\mbox{CS~22892--052}}
\def\gcsixeight{\mbox{M68}}

\def\apj{ApJ}
\def\aap{A\&A}
\def\aj{AJ}
\def\mnras{MNRAS}
\def\araa{ARA\&A}
\def\apjs{ApJS}
\def\apjl{ApJL}
\def\procspie{Proc.\ SPIE}
\def\jqsrt{J.\ Quant.\ Spectrosc.\ Rad.\ Trans.}
\def\pasp{PASP}
\def\pasj{PASJ}
\def\physscr{Phys.\ Scr.}

\title[Abundances in Segue 2]
{Detailed Abundance Analysis of the Brightest Star in Segue 2,
the Least Massive Galaxy\thanks{
This paper includes data gathered with the 6.5 meter 
Magellan Telescopes located at Las Campanas Observatory, Chile.}
}

\author[Ian U.\ Roederer and Evan N.\ Kirby]{%
Ian U.\ Roederer$^{1}$\thanks{E-mail: iur@umich.edu} 
and
Evan N.\ Kirby$^{2,3}$\thanks{E-mail: ekirby@uci.edu}\\
$^{1}$Department of Astronomy, University of Michigan,
500 Church Street, Ann Arbor, MI 48109, USA\\
$^{2}$Department of Physics and Astronomy, University of California,
4129 Frederick Reines Hall, Irvine, CA 92697, USA\\
$^{3}$Center for Galaxy Evolution Fellow
}

\begin{document}

\pagerange{\pageref{firstpage}--\pageref{lastpage}} 
\pubyear{2014}
\maketitle
\label{firstpage}

\begin{abstract}

We present the first high resolution spectroscopic observations of
one red giant star in the ultra-faint dwarf galaxy Segue~2, 
which has the lowest total mass 
(including dark matter) estimated for any known galaxy.
These observations were made using the 
MIKE spectrograph on the Magellan~II Telescope
at Las Campanas Observatory.
We perform a standard abundance analysis of this star,
SDSS~J021933.13$+$200830.2,
and present abundances of 21~species of 18~elements
as well as upper limits for 25~additional species.
We derive [Fe/H]~$= -$2.9, in excellent agreement with
previous estimates from medium resolution spectroscopy.
Our main result is that this star 
bears the chemical signatures
commonly found in field stars of similar metallicity.
The heavy elements produced by neutron-capture reactions
are present, but they are deficient at levels
characteristic of stars in other ultra-faint dwarf galaxies
and a few luminous dwarf galaxies.
The otherwise normal abundance patterns suggest
that the gas from which this star formed was enriched
by metals from multiple Type~II supernovae reflecting 
a relatively well-sampled IMF.
This adds to the growing body of evidence
indicating that Segue~2 may have been
substantially more massive in the past.

\end{abstract}


\section{Introduction}
\label{introduction}

In the last decade, we have witnessed the discovery of 
extremely low-luminosity dwarf galaxies, largely due to the
advent of the Sloan Digital Sky Survey (SDSS).  
The innovative
searches for these galaxies were pioneered by
\citet{willman05}, \citet{belokurov07}, and others.
One member of this class of ultra-faint dwarf galaxies,
\seggal, was initially identified by \citet{belokurov09}
as a stellar overdensity on the sky 
in images obtained as part of the 
Sloan Extension for Galactic Understanding and Exploration (SEGUE)
and by using matched filters in colour-magnitude space.
Deeper imaging and followup spectroscopy confirmed this detection,
identified the presence of a cold velocity structure
with a non-zero velocity dispersion,
and indicated a mean metallicity of approximately 1/100$^{\rm th}$
solar.

\citet{kirby13} used the DEIMOS spectrograph to obtain
red and near-infrared spectra of 25~probable members of \seggal.
These medium resolution spectra
allowed \citeauthor{kirby13}\ to 
constrain the line-of-sight velocity dispersion of \seggal\ 
to be $<$~2.6~\kmsec\ at 95~per cent confidence.
This corresponds to a mass $<$~2.1~$\times$~10$^{5}$~\msun\
within the half-light radius
assuming \seggal\ is in dynamical equilibrium.
That study also derived abundances of iron (Fe) and 
the $\alpha$ elements
magnesium (Mg), silicon (Si), calcium (Ca), and
titanium (Ti)
for the 10~brightest members of \seggal\
based on fits to a grid of synthetic spectra.
\citeauthor{kirby13}\ confirmed that stars in \seggal\
span a range in metallicity of more than 1.5~dex.
The [$\alpha$/Fe] ratios in \seggal\ generally decrease
with increasing [Fe/H], as seen in 
classical dwarf galaxies
(e.g., \citealt{shetrone03}; \citealt{kirby11a})
but not all of the ultra-faint dwarf galaxies
\citep{frebel10,vargas13}.
\seggal\ is the least-massive galaxy currently known
based on its inferred dynamical mass.
%The inferred mass of \seggal\ makes it the least-massive
%galaxy currently known (by dynamical--not stellar--mass).
Yet with a mean metallicity of
[Fe/H]~$= -$2.2 it does not obey the mass-metallicity
relationship established by \citet{kirby11b} for 
classical and ultra-faint dwarf galaxies.
\seggal, along with \segigal\ and \wilgal,
may reveal the existence of a metallicity floor in galaxy formation.
Alternatively,
\seggal\ may have been substantially more massive
before being tidally stripped down--by a factor of several
hundred in stellar mass--to the 
remnant observed today.

We present the first high resolution spectroscopic observations of
one red giant in \seggal, \seg, the only star
reasonably bright enough for such observations.
By coincidence, this star happens to be the most metal-poor star 
identified by \citet{kirby13} as a probable member.
We use these data to confirm the radial velocity measured by 
\citeauthor{kirby13}\
and derive detailed abundances of 21~species of 
18~elements.
We also present upper limits derived from non-detections 
of 25~additional species.
%While the abundance pattern of one star is of little
%utility when trying to understand the formation and evolution
%of an entire galaxy, we attempt to use these data to 
%identify sites of nucleosynthesis
%that occurred in \seggal\ prior to the formation of \seg.

Throughout this work we
adopt the standard definitions of elemental abundances and ratios.
For element X, the logarithmic abundance is defined
as the number of atoms of X per 10$^{12}$ hydrogen atoms,
$\log\epsilon$(X)~$\equiv \log_{10}(N_{\rm X}/N_{\rm H}) +$~12.0.
For elements X and Y, [X/Y] is 
the logarithmic abundance ratio relative to the solar ratio,
defined as $\log_{10} (N_{\rm X}/N_{\rm Y}) -
\log_{10} (N_{\rm X}/N_{\rm Y})_{\odot}$, using
like ionization states;
i.e., neutrals with neutrals and ions with ions.
We adopt the solar abundances listed in \citet{asplund09}.
Abundances or ratios denoted with the ionization state
indicate the total elemental abundance as derived from transitions of
that particular state.

\section{Observations}
\label{observations}

Only one star in \seggal, \seg, is bright enough for 
high-resolution spectroscopic observations.
Its $g$ magnitude, 17.18 
% $(g-r)_{0} =$~0.80
($V \approx$~16.60),
% $(B-V)_{0} \approx$~0.94),
is nevertheless quite faint 
when compared with
the majority of metal-poor 
field red giants in the solar neighborhood
that have comprised several large abundance surveys 
in recent decades.
Spectroscopic observations of similar quality
for other members of \seggal\
are not likely in the near future
because the next-brightest probable member of \seggal\ is more than 
two magnitudes fainter.

\begin{table}
\begin{minipage}{\textwidth}
\caption{Log of Observations
\label{obstab}}
%\tablewidth{0pt}
%\tabletypesize{\scriptsize}
\begin{tabular}{cccc}
\hline
Date &
UT mid- &
Exp. & 
Heliocentric \\
 &
exposure &
time (s) &
$V_{\rm r}$ (\kmsec) \\
\hline
2013 Dec 19 & 01:46 & 6600 & $-$39.5 \\
2013 Dec 20 & 01:47 & 6600 & $-$39.6 \\
2013 Dec 21 & 01:45 & 6600 & $-$40.0 \\
2013 Dec 22 & 01:49 & 4400 & $-$39.9 \\
2013 Dec 23 & 01:57 & 8000 & $-$38.9 \\
\hline
\end{tabular}
\end{minipage}
\end{table}

Table~\ref{obstab} presents a record of our observations of \seg.
Observations were made with the Magellan Inamori Kyocera Echelle (MIKE)
spectrograph \citep{bernstein03}
on the 
6.5~m Landon Clay Telescope (Magellan~II)
at Las Campanas Observatory.
These spectra were taken with the 0\farcs7\,$\times$\,5\farcs0 slit, 
yielding
a resolving power of $R \equiv \lambda/\Delta\lambda \sim$~41,000 in the blue 
and $R \sim$~35,000 in the red as measured from isolated ThAr lines
in the comparison lamp images.
This corresponds to $\approx$~2.5 and 2.1~pixels per resolution element
on the blue and red arms, respectively.
The blue and red arms are split by a dichroic at $\approx$~4950~\AA.
This setup achieves complete wavelength coverage from 
3350--9150~\AA, although
only the spectra longward of $\approx$~4000~\AA\ 
have signal sufficient to perform a detailed abundance analysis.
Data reduction,
extraction, sky subtraction,
and wavelength calibration were performed using 
the MIKE data reduction pipeline
written by D.\ Kelson
(see \citealt{kelson03}).
Coaddition and continuum normalization 
were performed within
\textsc{iraf}\@.

\begin{figure*}
\centering
\includegraphics[angle=270,width=5.0in]{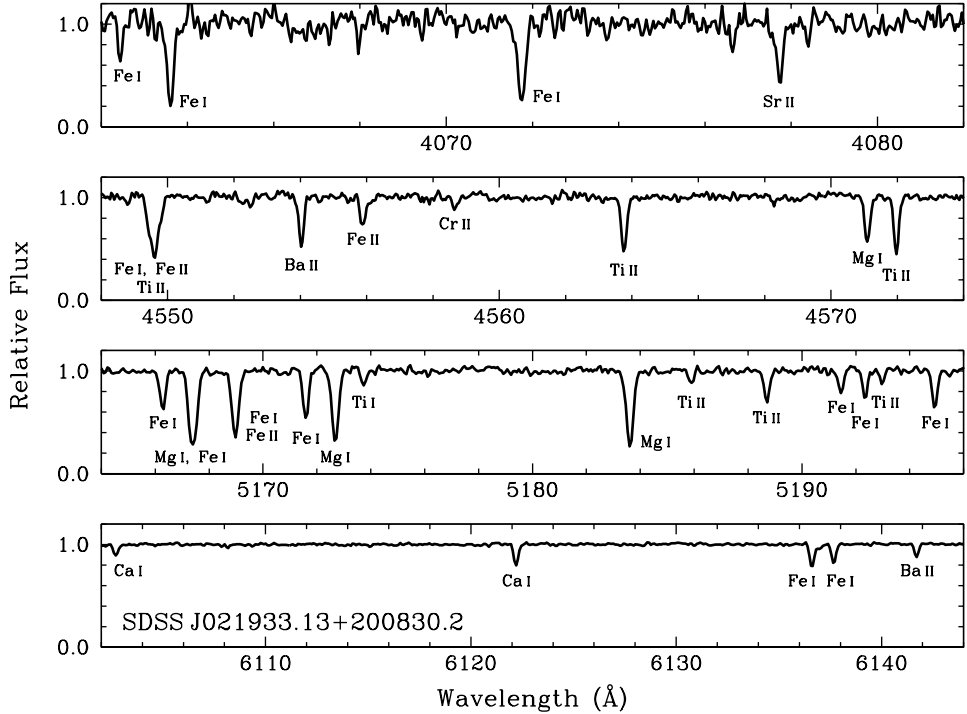}
\caption{
\label{spectplot}
Four regions of the spectrum of \segalt,
including several lines of interest.
The species responsible for each absorption line is indicated.
}
\end{figure*}

Fig.~\ref{spectplot} illustrates
four sample regions of our spectrum of \seg.
Several prominent absorption features are indicated.
After coaddition of the individual observations,
totaling 8.9~h of integration, our spectrum of \seg\ 
has signal to noise (S/N) levels of 20, 55, 60, and 145
per pixel in the continuum
at 3950, 4550, 5200, and 6750~\AA, respectively.
\seg\ is a northern hemisphere target, and
the differential refraction
caused by observing at zenith angles $>$~50$^{\circ}$
is apparent in the compromised signal at blue wavelengths.

\section{Radial Velocity Measurements}
\label{rv}

We measure radial velocities ($V_{r}$) for each observation
using the \textsc{iraf} \textit{fxcor} task to cross-correlate
several individual orders with prominent absorption lines.
The statistical uncertainties are 
smaller than 0.7~\kmsec\ per observation.
We establish the velocity zeropoint by cross-correlating our
highest S/N observation against a metal-poor velocity standard
star, \mbox{HD~128279}, 
acquired previously and analyzed by \citet{roederer14}.
We also obtained four high-S/N observations 
of the velocity standard star
\mbox{HD~83212},
for which we measure $V_{r} = +$109.3~$\pm$~0.2~\kmsec.
This agrees well with the mean velocity published
by \citet{carney03},
$V_{r} = +$109.13~$\pm$~0.16,
indicating that our velocity zeropoint is reliable at the
$\approx$~0.2~\kmsec\ level.
We measure a mean Heliocentric radial velocity for \seg\ 
of $V_{r} = -$39.6~$\pm$~0.2~\kmsec.
This is in excellent agreement with the value measured by \citet{kirby13}
from moderate resolution DEIMOS spectra,
$-$41.0~$\pm$~2.0~\kmsec.

\section{Model Atmosphere Parameters}
\label{modelatm}

We use one-dimensional model atmospheres interpolated from the
\textsc{atlas9} grid \citep{castelli03}.
We perform the analysis using a recent version of the spectral
line analysis code \textsc{moog} 
(\citealt{sneden73}; see discussion in \citealt{sobeck11}).
Both \textsc{atlas9} and 
\textsc{moog} assume that local thermodynamic equilibrium
(LTE) holds in the line-forming layers.

\begin{figure}
\centering
\includegraphics[angle=00,width=3.0in]{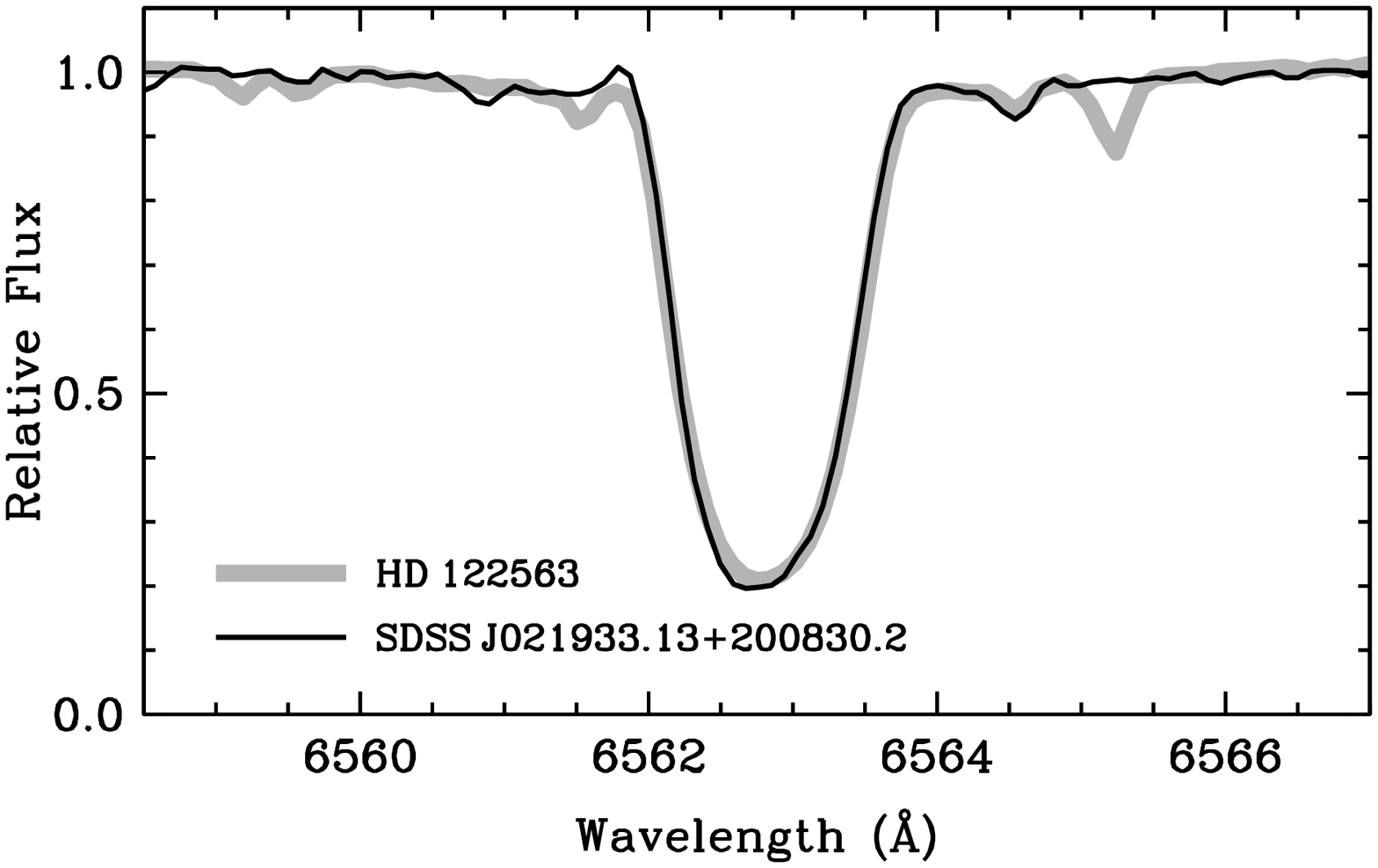}
\caption{
\label{halphaplot}
Profile of the H~$\alpha$ line in 
\segalt\ and \mbox{HD~122563}.
The spectrum of \mbox{HD~122563} was obtained
using the same MIKE setup and presented by \citet{roederer14}.
}
\end{figure}

We make several estimates of the effective temperature (\teff) of \seg.
\citet{ivezic08} presented a
colour-\teff\ relationship tailored for
SDSS broadband photometry; using the 
dereddened $g-r$ colour, their equation 3
predicts 4588~K.
The set of Padova 12.6~Gyr isochrones \citep{girardi02,bressan12}
predicts 4597~K from the dereddened $g-i$ colour.
\citet{kirby13} derived 4566~K from the DEIMOS spectrum
based on a combination of photometry and spectroscopy.
Following the techniques described by \citet{kirby08} and
\citet{kirby09,kirby10},
a $\chi^2$ minimization algorithm determines which
synthetic spectrum best matches the observed one.
The temperature is allowed to vary but is
constrained to a range of about 200~K around the 
photometrically-determined \teff.
These three estimates of \teff\ agree well.
Deriving \teff\ by a more traditional spectroscopic approach,
requiring that the iron abundances derived from Fe~\textsc{i} lines
show no trend with excitation potential, would demand
a \teff\ lower by nearly 350~K.
In Fig.~\ref{halphaplot}, we illustrate the region
surrounding the H~$\alpha$ line in \seg\ and \hd,
a metal-poor giant
with \teff~$=$~4598~$\pm$~41~K
\citep{creevey12}.
The line profiles are indistinguishable, suggesting
the warmer temperatures are appropriate for \seg.
For consistency with \citet{kirby13},
we adopt \teff~$=$~4566~K.

We calculate a physical surface gravity, \logg,
based on standard stellar relations and the distance
to \seggal\ estimated by the
apparent magnitude of four blue horizontal branch stars
\citep{belokurov09}.
Other distance estimates by \citet{ripepi12}--who
compared the \seggal\ fiducial sequence with 
that of globular cluster \gcsixeight, and 
\citet{boettcher13}--who characterized the pulsation
cycle of an RR~Lyrae star in \seggal, are
consistent with the \citeauthor{belokurov09}\ value.
For a distance modulus of $m-M =$~17.7~$\pm$~0.1,
our adopted \teff~$=$~4566~K,
\teff$_{\odot} =$~5777~K, 
\logg$_{\odot} =$~4.44,
$M_{\rm bol \odot} =$~4.74,
$g_{0} =$~16.46,
a bolometric correction, $BC_{g} = -$1.05 \citep{girardi04},
and a mass of 0.72~M$_{\odot}$ (from the 
12.6~Gyr isochrones of \citealt{bressan12}),
we calculate \logg~$=$~1.36.
Reasonable uncertainties on these quantities
imply an uncertainty in \logg\ of no more than
0.10--0.15~dex.

We derive the microturbulence velocity, \vt, 
by requiring that the abundances derived from Fe~\textsc{i} lines
show no correlation with the reduced equivalent width,
\logrw.
Our derived value, 1.85~\kmsec, is in excellent agreement
with the value calculated by \citet{kirby13}, 1.87~\kmsec,
using a relationship between \logg\ and \vt.

We set the overall model metallicity, [M/H], 
equal to the iron abundance derived from Fe~\textsc{ii} lines.
We iteratively cull iron lines deviating by more than 
2~$\sigma$ from the mean.
The adopted model metallicity, [Fe~\textsc{ii}/H]~$= -$2.81, is slightly
higher than that derived from Fe~\textsc{i} lines,
[Fe~\textsc{i}/H]~$= -$2.96.
We would need to lower \logg\ by 0.45~dex to
force these values to be equivalent, which seems unlikely
based on the discussion above.
This difference is what might be expected
if the non-LTE effect of iron overionization 
is occurring (e.g., \citealt{thevenin99}).
This difference is within the range found
for cool metal-poor giants by \citet{roederer14}.
If we were to adopt the cooler \teff\ suggested by the
iron excitation equilibrium method without
adjusting \logg\ accordingly, the difference between 
the iron abundances derived from Fe~\textsc{i} and \textsc{ii}
would be nearly 0.8~dex.
In order to compensate, we would have lowered \vt\ by 0.4~\kmsec,
which is in the opposite sense than would be expected if
\seg\ is a more evolved giant.
We conclude by noting that our derived metallicity is in
good agreement with what \citet{kirby13} found by
fitting numerous iron lines in the red part of the spectrum,
[Fe/H]~$= -$2.85~$\pm$~0.11.

\citet{roederer14} compared their derived model atmosphere parameters
with those derived by a variety of different methods
for stars in common.
For red giants, the residuals had standard deviations of
151~K in \teff,
0.40 in \logg,
0.41~\kmsec\ in \vt, and
0.24~dex in [Fe~\textsc{ii}/H].
For \logg\ we retain the $\approx$~0.15~dex uncertainty
estimated previously, but for \teff, \vt, and [M/H]
we adopt these values as the systematic uncertainty in 
the model atmosphere parameters.

\section{Abundance Analysis}
\label{analysis}

Table~\ref{atomictab} lists the
line wavelength, species identification,
excitation potential (E.P.)\ of the lower electronic energy level,
and \loggf\ value
for each transition examined.
Equivalent widths (EW), also listed in Table~\ref{atomictab}, 
are measured using a
semi-automatic routine that fits Voigt absorption line profiles
to continuum-normalized spectra
\citep{roederer14}.
When a line is not detected, 
we derive 3~$\sigma$ upper limits on the abundance using
a version of the formula presented on p.\ 590 of
\citet{frebel08}, which itself is derived from
equation A8 of \citet{bohlin83}.
When multiple lines of the same species are not detected,
we adopt the upper limit that provides the
strongest constraint on the abundance.

\begin{table*}
\begin{minipage}{\textwidth}
\caption{Atomic Data, Equivalent Widths, and Derived Abundances
\label{atomictab}}
%\tablewidth{0pt}
%\tabletypesize{\scriptsize}
\begin{tabular}{cccccccc}
\hline
Species &
$\lambda$ &
E.P.\ &
$\log gf$ &
Ref. &
EW\footnote{Entries with no EW listed indicate that
no line is detected or
spectrum synthesis matching is used to derive an abundance.} &
$\log \epsilon$ &
$\sigma$ \\
  & 
 (\AA) &
 (eV) & 
  & 
  & 
 (m\AA) &
  &
  \\
\hline
    Li~\textsc{i}  & 6707.80 & 0.00 & $+$0.17 &  1 & \ldots  & $<$ 0.28 & \ldots  \\
C (CH) & \multicolumn{3}{c}{$A^2\Delta - X^2\Pi$ G band}        &  2 & \ldots  &     4.85 &   0.20  \\
N (CN) & \multicolumn{3}{c}{$B^2\Sigma - X^2\Sigma$ violet band}&  3 & \ldots  & $<$ 6.85 & \ldots  \\
N (CN) & \multicolumn{3}{c}{$A^2\Sigma - X^2\Sigma$ red band}   &  3 & \ldots  & $<$ 9.00 & \ldots  \\
  ~[O~\textsc{i}]  & 6300.30 & 0.00 & $-$9.78 &  4 & \ldots  & $<$ 6.74 & \ldots  \\
     O~\textsc{i}  & 7771.94 & 9.15 & $+$0.37 &  4 & \ldots  & $<$ 6.68 & \ldots  \\
     O~\textsc{i}  & 7774.17 & 9.15 & $+$0.22 &  4 & \ldots  & $<$ 6.82 & \ldots  \\
     O~\textsc{i}  & 7775.39 & 9.15 & $+$0.00 &  4 & \ldots  & $<$ 7.16 & \ldots  \\
    Na~\textsc{i}  & 5889.95 & 0.00 & $+$0.11 &  4 &  189.2  &     3.45 &    0.49 \\
    Na~\textsc{i}  & 5895.92 & 0.00 & $-$0.19 &  4 &  161.1  &     3.42 &    0.53 \\
\hline
\end{tabular}
\\
The complete version of Table~\ref{atomictab} is 
available online.
An abbreviated version is shown here to demonstrate its form and content.
References:\ 
 (1) \citealt{smith98} for both \loggf\ value and $^{7}$Li hfs;
 (2) B.\ Plez 2007, private communication;
 (3) \citealt{kurucz95};
 (4) \citealt{fuhr09};
 (5) \citealt{chang90};
 (6) \citealt{aldenius07};
 (7) \citealt{aldenius09};
 (8) \citealt{lawler89}, using hfs from \citealt{kurucz95};
 (9) \citealt{lawler13};
(10) \citealt{pickering01}, with corrections given in \citealt{pickering02};
(11) \citealt{wood13};
(12) \citealt{doerr85a}, using hfs from \citealt{kurucz95};
(13) \citealt{biemont89};
(14) \citealt{sobeck07};
(15) \citealt{nilsson06};
(16) \citealt{denhartog11} for both \loggf\ value and hfs;
(17) \citealt{obrian91};
(18) \citealt{nitz99}, using hfs from \citealt{kurucz95};
(19) \citealt{fuhr09}, using hfs from \citealt{kurucz95};
(20) \citealt{roederer12a};
(21) \citealt{biemont11};
(22) \citealt{ljung06};
(23) \citealt{palmeri05};
(24) \citealt{fuhr09}, using hfs/IS from \citealt{mcwilliam98} when available;
(25) \citealt{lawler01la}, using hfs from \citealt{ivans06};
(26) \citealt{lawler09}; 
(27) \citealt{li07}, using hfs from \citealt{sneden09};
(28) \citealt{denhartog03}, using hfs/IS from \citealt{roederer08} 
        when available;
(29) \citealt{lawler06}, using hfs/IS from \citealt{roederer08} when available;
(30) \citealt{lawler01eu}, using hfs/IS from \citealt{ivans06};
(31) \citealt{denhartog06};
(32) \citealt{roederer12b};
(33) \citealt{lawler01tb}, using hfs from \citealt{lawler01tbhfs} 
        when available;
(34) \citealt{wickliffe00}; 
(35) \citealt{lawler04} for both \loggf\ value and hfs;
(36) \citealt{lawler08};
(37) \citealt{wickliffe97};
(38) \citealt{sneden09} for both \loggf\ value and hfs/IS;
(39) \citealt{lawler07};
(40) \citealt{ivarsson03}, using hfs/IS from \citealt{cowan05}--see
       note on \loggf\ values there;
(41) \citealt{biemont00}, using hfs/IS from \citealt{roederer12b}. \\
\end{minipage}
\end{table*}

Our abundance analysis closely follows that
of \citet{roederer14}.
Spectrum synthesis matching is performed for lines broadened by
hyperfine splitting (hfs) or in cases where
a significant isotope shift (IS) may be present.
Linelists are generated using the \citet{kurucz95} lists
and updated using more recent experimental data when available.
For unblended lines we use \textsc{moog} to compute
theoretical EWs, which are then forced to match
measured EWs by adjusting the abundance.
For heavy elements with multiple isotopes,
we adopt the \rpro\ isotopic ratios 
presented in \citet{sneden08}.
We examine the stellar spectrum simultaneously with
a spectrum of earth's atmosphere \citep{hinkle00}
for all lines with $\lambda >$~5670~\AA.
We do not attempt to derive abundances from any of these lines
when the telluric spectrum suggests they may be compromised.
Carbon abundances, derived from the CH G band near 4300~\AA,
and nitrogen upper limits, derived from the violet and red CN bands near
3880 and 8000~\AA\ are also
found via spectrum synthesis matching.

Table~\ref{atomictab} lists the abundances
derived from each of the 260~features examined.
We adopt corrections
to account for departures from LTE
for Na~\textsc{i} lines \citep{lind11} and
K~\textsc{i} lines \citep{takeda02}.
These corrections amount to 
$-$0.35, $-$0.33, and $-$0.31~dex, respectively,
for the Na~\textsc{i} 5889, 5895, and 
K~\textsc{i} 7698~\AA\ lines.
We also adopt the line-by-line corrections 
derived by \citet{roederer14} to account for 
which lines are used in the analysis;
these corrections are listed in Table~16 of that study.

Table~\ref{abundtab} lists the mean abundances
derived for \seg.
Weighted mean abundances and uncertainties are computed using the
formalism presented in \citet{mcwilliam95},
as discussed in detail in \citeauthor{roederer14} %.
We remind readers that the [X/Fe] ratios are constructed
by referencing abundances of element X,
derived from neutral (ionized) species, to the 
iron abundance derived from Fe~\textsc{i} (Fe~\textsc{ii}) lines.
Several sets of uncertainties are listed in Table~\ref{abundtab}.
The statistical uncertainty, $\sigma_{\rm statistical}$,
accounts for uncertainties in the EWs, \loggf\ values,
non-LTE corrections, and line-by-line offset
corrections.
The total uncertainty, $\sigma_{\rm total}$,
accounts for the statistical uncertainty and
uncertainties in the model atmosphere parameters.
The other two uncertainties listed in
Table~\ref{abundtab}
are approximations to the abundance ratio uncertainties
given by equations~A19 and A20 of \citeauthor{mcwilliam95} %.
The quantity $\sigma_{\rm neutrals}$ for element A
should be added in quadrature with $\sigma_{\rm statistical}$ for
element B when computing the ratio [A/B] when B is
derived from neutral lines.
Similarly, $\sigma_{\rm ions}$ for element A
should be added in quadrature with $\sigma_{\rm statistical}$ for
element B when element B is derived from ionized lines.

\begin{table*}
\begin{minipage}{\textwidth}
\caption{Mean Abundances in SDSS J021933.13$+$200830.2
\label{abundtab}}
%\tablewidth{0pt}
%\tabletypesize{\scriptsize}
%\tabletypesize{\tiny}
\begin{tabular}{cccccccc}
\hline
Species &
$N_{\rm lines}$ &
$\log\epsilon$ &
[X/Fe]\footnote{[Fe/H] is indicated for Fe~\textsc{i} and Fe~\textsc{ii}} &
$\sigma_{\rm statistical}$ &
$\sigma_{\rm total}$ &
$\sigma_{\rm neutrals}$ &
$\sigma_{\rm ions}$ \\
\hline
 Fe~\textsc{i}  &  96 &     4.54 & $-$2.96 &  0.03 &  0.19 &  0.00 &  0.00  \\
 Fe~\textsc{ii} &  10 &     4.69 & $-$2.81 &  0.05 &  0.08 &  0.00 &  0.00  \\
 Li~\textsc{i}  &   1 & $<$ 0.28 & \ldots  & \ldots& \ldots& \ldots& \ldots \\
 C~(CH)         &   1 &     4.85 & $-$0.52 & \ldots& \ldots& \ldots& \ldots \\
 N~(CN)         &   1 & $<$ 6.85 &$<$ 1.81 & \ldots& \ldots& \ldots& \ldots \\
 O~\textsc{i}   &   4 & $<$ 6.68 &$<$ 0.95 & \ldots& \ldots& \ldots& \ldots \\
 Na~\textsc{i}  &   2 &     3.44 &    0.16 &  0.10 &  0.40 &  0.29 &  0.37  \\
 Mg~\textsc{i}  &   5 &     4.95 &    0.31 &  0.03 &  0.21 &  0.05 &  0.18  \\
 Al~\textsc{i}  &   1 &     2.89 & $-$0.60 &  0.07 &  0.35 &  0.23 &  0.31  \\
 Si~\textsc{i}  &   1 &     4.70 &    0.15 &  0.15 &  0.34 &  0.21 &  0.30  \\
 K~\textsc{i}   &   1 &     2.37 &    0.30 &  0.12 &  0.22 &  0.13 &  0.20  \\
 Ca~\textsc{i}  &  12 &     3.54 &    0.17 &  0.10 &  0.21 &  0.11 &  0.19  \\
 Sc~\textsc{ii} &   6 &     0.53 &    0.19 &  0.03 &  0.07 &  0.16 &  0.06  \\
 Ti~\textsc{i}  &  13 &     2.00 &    0.01 &  0.02 &  0.18 &  0.05 &  0.16  \\
 Ti~\textsc{ii} &  25 &     2.42 &    0.28 &  0.03 &  0.07 &  0.16 &  0.06  \\
 V~\textsc{i}   &   1 &     0.85 & $-$0.12 &  0.10 &  0.21 &  0.11 &  0.18  \\
 V~\textsc{ii}  &   2 & $<$ 1.10 &$<-$0.02 & \ldots& \ldots& \ldots& \ldots \\
 Cr~\textsc{i}  &   7 &     2.39 & $-$0.28 &  0.02 &  0.18 &  0.05 &  0.16  \\
 Cr~\textsc{ii} &   2 &     3.01 &    0.18 &  0.04 &  0.08 &  0.17 &  0.07  \\
 Mn~\textsc{i}  &   3 &     1.96 & $-$0.51 &  0.02 &  0.18 &  0.05 &  0.16  \\
 Co~\textsc{i}  &   1 &     1.64 & $-$0.39 &  0.10 &  0.27 &  0.15 &  0.24  \\
 Ni~\textsc{i}  &   4 &     3.33 &    0.08 &  0.10 &  0.21 &  0.11 &  0.19  \\
 Cu~\textsc{i}  &   1 & $<$ 0.88 &$<-$0.35 & \ldots& \ldots& \ldots& \ldots \\
 Zn~\textsc{i}  &   2 &     1.97 &    0.37 &  0.02 &  0.18 &  0.05 &  0.16  \\
 Rb~\textsc{i}  &   1 & $<$ 1.24 &$<$ 1.68 & \ldots& \ldots& \ldots& \ldots \\
 Sr~\textsc{ii} &   2 &  $-$1.29 & $-$1.35 &  0.02 &  0.29 &  0.27 &  0.26  \\
 Y~\textsc{ii}  &   2 & $<-$1.53 &$<-$0.93 & \ldots& \ldots& \ldots& \ldots \\
 Zr~\textsc{ii} &   3 & $<-$0.74 &$<-$0.51 & \ldots& \ldots& \ldots& \ldots \\
 Tc~\textsc{i}  &   1 & $<-$0.04 & \ldots  & \ldots& \ldots& \ldots& \ldots \\
 Ba~\textsc{ii} &   5 &  $-$1.62 & $-$1.00 &  0.04 &  0.08 &  0.16 &  0.06  \\
 La~\textsc{ii} &   4 & $<-$1.74 &$<-$0.03 & \ldots& \ldots& \ldots& \ldots \\
 Ce~\textsc{ii} &   5 & $<-$1.30 &$<-$0.07 & \ldots& \ldots& \ldots& \ldots \\
 Pr~\textsc{ii} &   4 & $<-$1.59 &$<$ 0.50 & \ldots& \ldots& \ldots& \ldots \\
 Nd~\textsc{ii} &   4 & $<-$1.32 &$<$ 0.07 & \ldots& \ldots& \ldots& \ldots \\
 Sm~\textsc{ii} &   1 & $<-$1.41 &$<$ 0.44 & \ldots& \ldots& \ldots& \ldots \\
 Eu~\textsc{ii} &   4 & $<-$2.59 &$<-$0.30 & \ldots& \ldots& \ldots& \ldots \\
 Gd~\textsc{ii} &   3 & $<-$0.79 &$<$ 0.95 & \ldots& \ldots& \ldots& \ldots \\
 Tb~\textsc{ii} &   2 & $<-$1.34 &$<$ 1.17 & \ldots& \ldots& \ldots& \ldots \\
 Dy~\textsc{ii} &   3 & $<-$1.30 &$<$ 0.41 & \ldots& \ldots& \ldots& \ldots \\
 Ho~\textsc{ii} &   3 & $<-$1.61 &$<$ 0.72 & \ldots& \ldots& \ldots& \ldots \\
 Er~\textsc{ii} &   3 & $<-$0.78 &$<$ 1.11 & \ldots& \ldots& \ldots& \ldots \\
 Tm~\textsc{ii} &   2 & $<-$1.38 &$<$ 1.33 & \ldots& \ldots& \ldots& \ldots \\
 Yb~\textsc{ii} &   1 & $<-$0.76 &$<$ 1.13 & \ldots& \ldots& \ldots& \ldots \\
 Hf~\textsc{ii} &   2 & $<-$0.47 &$<$ 1.49 & \ldots& \ldots& \ldots& \ldots \\
 Ir~\textsc{i}  &   1 & $<$ 0.08 &$<$ 1.66 & \ldots& \ldots& \ldots& \ldots \\
 Pb~\textsc{i}  &   1 & $<$ 0.93 &$<$ 1.85 & \ldots& \ldots& \ldots& \ldots \\
\hline
\end{tabular}
\end{minipage}
\end{table*}

We note that comparisons with the abundance ratios derived
by \citet{kirby13} are not particularly meaningful since
the lines used in the present analysis are 
generally not covered by the DEIMOS spectrum.
In principle this should not matter, but in practice 
the small number of weak
Mg~\textsc{i} and Si~\textsc{i} lines that probably contribute most to
the \citeauthor{kirby13}\ results are all high excitation lines
that are quite sensitive to the adopted value of \teff\ 
in the model atmosphere.
Noise may also be an issue when only a few features are present,
as is the case in low metallicity stars.
These effects may contribute to the different [Mg/Fe] ratios 
($+$0.76~$\pm$~0.25 
found by \citeauthor{kirby13};
 $+$0.31~$\pm$~0.21 found here).
The silicon abundances derived in our study may be unreliable
(see Section~\ref{results}), so we do not make a
comparison with the [Si/Fe] ratio derived by \citeauthor{kirby13} %.
The other ratios 
presented in Table~4 of \citeauthor{kirby13}\
are in satisfactory agreement with those derived in the present study:\
for [Fe~\textsc{i}/H],  $-$2.85~$\pm$~0.11 
and $-$2.96~$\pm$~0.19, respectively;
for [Ca/Fe], $+$0.30~$\pm$~0.13 
and $+$0.17~$\pm$~0.21; and
for [Ti/Fe], $-$0.18~$\pm$~0.31 
and $+$0.01~$\pm$~0.18.

\section{Results}
\label{results}

\begin{figure*}
\centering
\includegraphics[angle=00,width=5.0in]{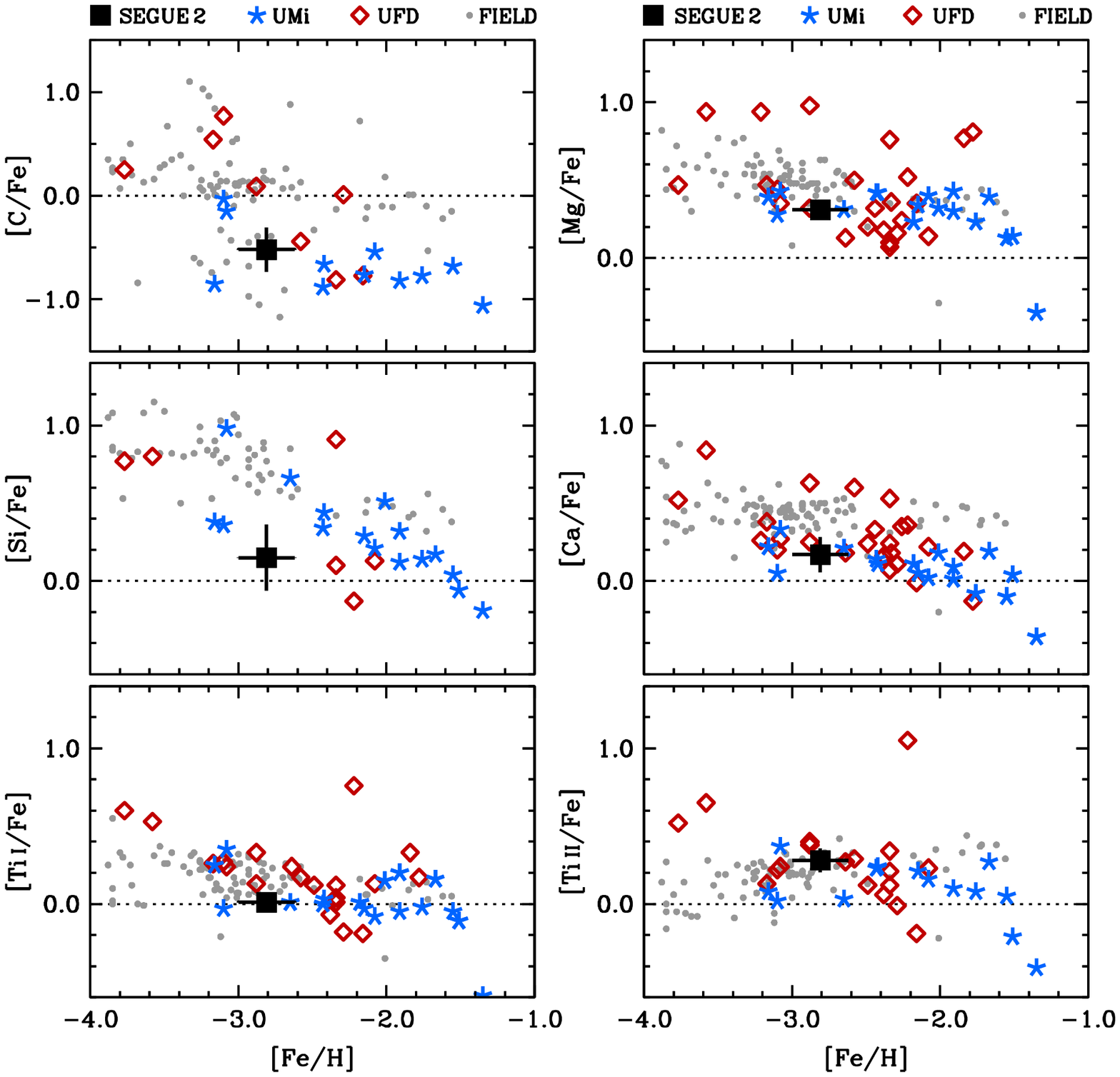}
\caption{
\label{abundplot1}
The [C/Fe], [Mg/Fe], [Si/Fe], [Ca/Fe], [Ti~\textsc{i}/Fe], and 
[Ti~\textsc{ii}/Fe] ratios in 
\segalt\ (``Segue~2''),
the classical dwarf spheroidal galaxy \umigal\ (``UMi''),
several ultra-faint dwarf galaxies (``UFD'';
\boogal, \comgal, \hergal, \leogal, \segigal, \umagal)
%Bo\"{o}tes~I,
%Coma Berenices,
%Hercules,
%Leo~IV,
%Segue~1, and
%Ursa Major~II)
and a sample of field red giants.
References for the comparison samples are listed in Section~\ref{results}.
The dotted lines represent the Solar ratios.
}
\end{figure*}

\begin{figure}
\centering
\includegraphics[angle=00,width=3.0in]{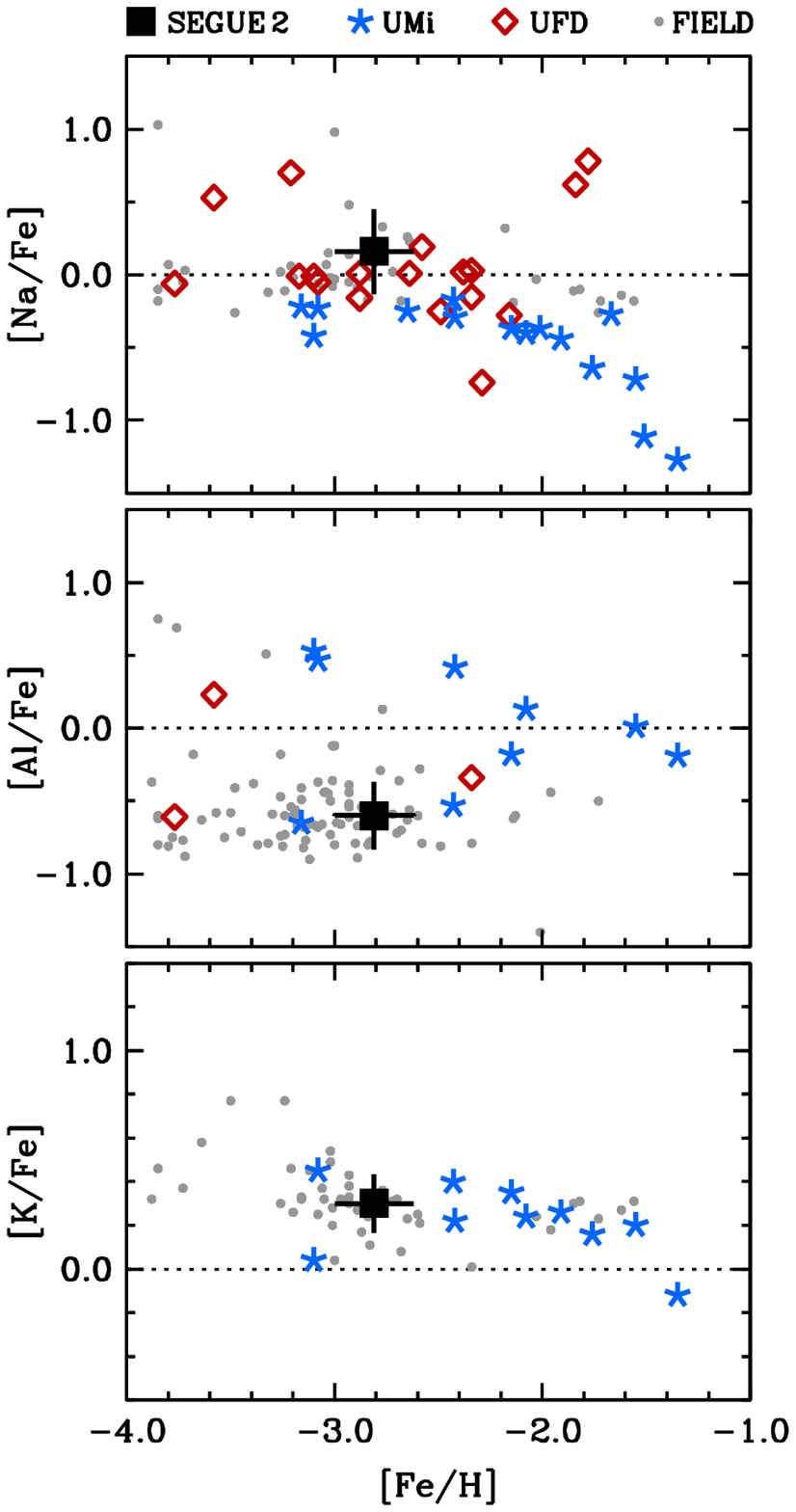}
\caption{
\label{abundplot2}
The [Na/Fe], [Al/Fe], and [K/Fe] ratios.
Symbols are the same as in Fig.~\ref{abundplot1}.
}
\end{figure}

\begin{figure*}
\centering
\includegraphics[angle=00,width=5.0in]{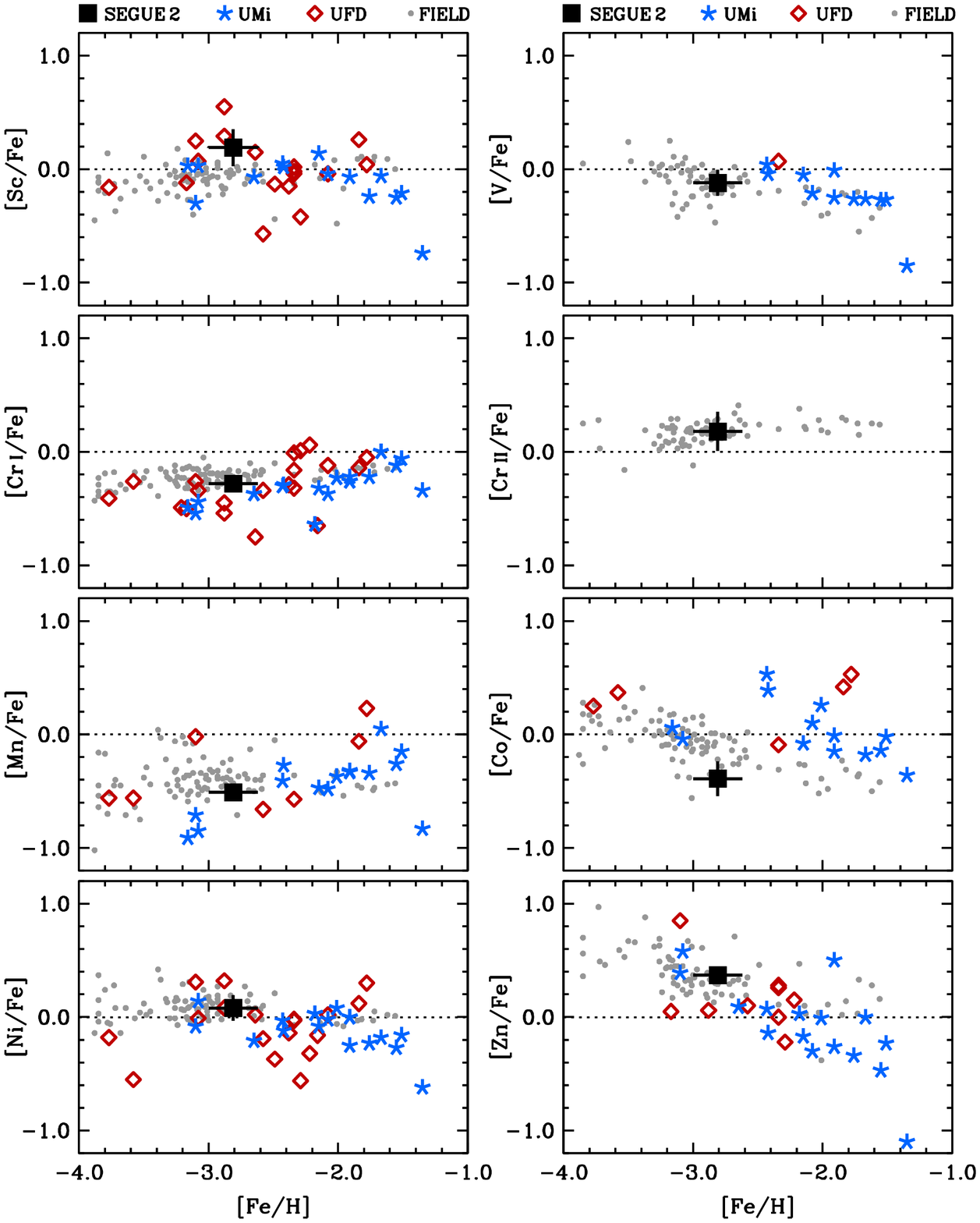}
\caption{
\label{abundplot3}
The [Sc/Fe], [V/Fe], [Cr~\textsc{i}/Fe], [Cr~\textsc{ii}/Fe],
[Mn/Fe], [Co/Fe], [Ni/Fe], and [Zn/Fe] ratios.
Symbols are the same as in Fig.~\ref{abundplot1}.
}
\end{figure*}

\begin{figure}
\centering
\includegraphics[angle=00,width=3.0in]{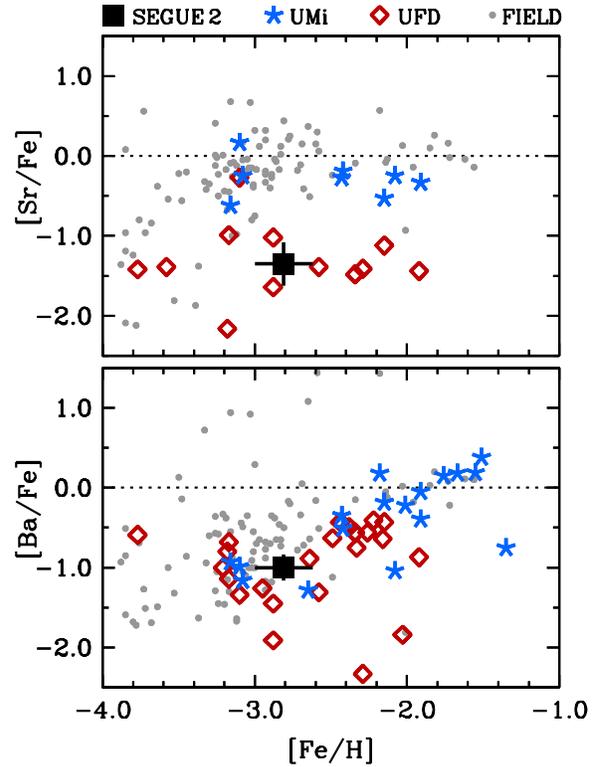}
\caption{
\label{abundplot4}
The [Sr/Fe] and [Ba/Fe] ratios.
Symbols are the same as in Fig.~\ref{abundplot1}.
}
\end{figure}

Figures~\ref{abundplot1} through \ref{abundplot5} 
illustrate our derived abundance ratios for \seg.
Several other sets of abundances derived
from high resolution spectroscopy are shown for comparison.
\citet{kirby13} noted the similarity between the
mean metallicity and 
abundance trends in \seggal\ and the more luminous
classical dwarf spheroidal galaxy \umigal,
which they propose as a present-day analog of
what \seggal\ once may have been in the tidal stripping scenario.
We illustrate abundances in \umigal\ using blue star symbols
in Figures~\ref{abundplot1} through \ref{abundplot5},
These data are compiled from the high resolution observations of
\citet{shetrone01},
\citet{sadakane04},
\citet{cohen10}, and 
\citet{kirby12}.
Abundances in several ultra-faint dwarf galaxies 
(\boogal, \comgal, \hergal, \leogal, \segigal, \umagal)
are marked by the red diamonds.
These data are compiled from 
\citet{koch08,koch13},
\citet{feltzing09},
\citet{frebel10},
\citet{norris10a,norris10b},
\citet{simon10}, 
\citet{gilmore13}, and
\citet{ishigaki14}.
We only illustrate each star once
if it has been analyzed by multiple investigators, and we
give preference to the highest quality observations.
%We have made no attempt to correct the literature abundances
%shown in these figures to a consistent \loggf\ or
%solar abundance scale, to account for which lines
%are used in each analysis, or to apply non-LTE corrections consistently.
%Such differences will generally be small relative
%to the ranges and trends under consideration here.
Figures~\ref{abundplot1} through \ref{abundplot5} 
also include abundances from a sample of 98~metal-poor 
field red giant stars analyzed by \citet{roederer14}.

As seen in Fig.~\ref{abundplot1},
the subsolar [C/Fe] ratio in \seg\ is typical for red giants 
that have experienced first dredge-up,
which mixes products of the CN-cycle to the surface.
Similar [C/Fe] ratios are found in stars belonging to
the the ultra-faint dwarf galaxies and \umigal.
Unfortunately, our upper limit on the nitrogen abundance
is uninteresting and only constrains
[N/Fe]~$< +$1.81.
Our upper limit on the oxygen abundance
is also uninteresting,
[O/Fe]~$< +$0.95.
Lithium is not detected in \seg, 
but the upper limit we derive is similar to the upper limits 
commonly found in field red giants.

Fig.~\ref{abundplot1} also illustrates the ratios of
$\alpha$ elements to iron.
The [Mg/Fe] ratio in \seg\ 
is within one standard deviation of the field giants
and falls within the range of [Mg/Fe] ratios found
in \umigal\ and the ultra-faint dwarf galaxies.
This result is unchanged if other comparison samples of
field stars are considered instead
(e.g., \citealt{cayrel04,honda04,lai08,yong13,cohen13}).
The [Si/Fe] ratio in \seg\ is lower by $\approx$~0.6~dex than the
[Si/Fe] ratios in field giants at similar metallicity.
This abundance is derived from only one strong line,
Si~\textsc{i} 3905~\AA, and the abundances inferred from this line 
are known to anti-correlate with \teff\
(e.g., \citealt{preston06}).
This effect is minimized somewhat by our use of only the
red giants in the \citet{roederer14} sample for comparison,
but fig.~27 of that study demonstrates that there is still a
slight trend with \teff\ even with this restriction.
We discard the [Si/Fe] ratio from further consideration.
The [Ca/Fe] ratio in \seg\ is on the low end of the
distribution found in field stars of similar metallicity,
typically at the 1--2~$\sigma$ level depending on which
comparison sample is used.
Given the uncertainty in the [Ca/Fe] ratio for \seg, $\pm$~0.11~dex,
and the typical dispersions in the [Ca/Fe] ratio for
the comparison samples ($\approx$~0.10--0.15~dex for the
samples listed above)
this discrepancy is probably not significant.
The [Ca/Fe] ratio in \seg\ is within the range found for
\umigal\ and the ultra-faint dwarf galaxies.

[Ti~\textsc{i}/Fe] and [Ti~\textsc{ii}/Fe]
are different by 0.26~dex, with [Ti~\textsc{ii}/Fe] being higher in \seg.
\citet{bergemann11} calculates that the non-LTE corrections
for [Ti~\textsc{i}/Fe] are $\approx +$0.2--0.3~dex
greater than for [Ti~\textsc{ii}/Fe] in \hd,
which has similar stellar parameters to \seg,
and these corrections would bring 
[Ti~\textsc{i}/Fe] and [Ti~\textsc{ii}/Fe] into agreement.
\citet{roederer14} identified a trend of increasing
[Ti~\textsc{ii}/Ti~\textsc{i}] with increasing metallicity
(see fig.~54 there),
which could be attributed to non-LTE effects.
At [Fe/H]~$= -$2.9, that study found 
[Ti~\textsc{ii}/Ti~\textsc{i}]~$\approx +$0.1, ranging from
$-$0.1 to $+$0.3~dex,
which includes our derived value for \seg.
[Ti~\textsc{i}/Fe] is
on the low side of the distribution for field giants
with similar metallicities;
[Ti~\textsc{ii}/Fe], however, is well within the normal
range for field giants.
For the atmospheric conditions found in \seg,
ground and low-lying levels of Ti~\textsc{ii} constitute the main
reservoir of titanium atoms,
so these levels cannot be significantly out of equilibrium
\citep{lawler13}.
When comparing ratios of one $\alpha$ element to another, 
however, the titanium abundance derived from Ti~\textsc{i}
lines may be preferable since the other $\alpha$ element abundances
are also derived from transitions of the neutral species.
Regardless of whether [Ti~\textsc{i}/Fe] or [Ti~\textsc{ii}/Fe]
more accurately represents the [Ti/Fe] ratio in \seg,
both are within the range found in \umigal\ and the
ultra-faint dwarfs.

Fig.~\ref{abundplot2} illustrates the [Na/Fe], [Al/Fe], and [K/Fe] ratios.
%Corrections for departures for LTE have been applied 
%for Na~\textsc{i} and K~\textsc{i} lines.
These three abundance ratios in \seg\ all coincide with
the bulk of the same ratios in the field giants, 
Ursa Minor, and for [Na/Fe] and [Al/Fe] in the
ultra-faint dwarf galaxies.
Potassium has not been detected previously in 
any star in an ultra-faint dwarf galaxy,
and the [K/Fe] ratio in this one star in \seggal\ 
appears quite normal in comparison with the field giants
and stars in \umigal.

Fig.~\ref{abundplot3} illustrates the abundance ratios
among the iron group elements in \seg\ and the comparison samples.
In nearly all cases, the \seg\ ratios fall well within the 
ranges found in the comparison samples at similar metallicities.
The [Sc/Fe] ratio in \seg\ is $\approx$~0.2~dex higher than the
field giants, but it is within the range of the stars
in ultra-faint dwarf galaxies.
The [Co/Fe] ratio in \seg\ is lower than the limited
data available for stars in the ultra-faint dwarf galaxies,
but it is within the range of field giants.
These mild differences are probably not significant,
and we conclude that the ratios among the iron group
elements in \seg\ are normal for 
stars of this metallicity.

Fig.~\ref{abundplot4} illustrates the [Sr/Fe] and [Ba/Fe] ratios.
Both ratios are deficient in \seg\ by more than an order of magnitude 
relative to the solar ratios.
The [Sr/Fe] ratio appears normal with respect to the
ultra-faint dwarf galaxies, but it is noticeably lower than
the [Sr/Fe] ratios found in \umigal\ or the 
field stars at similar metallicity.
The [Ba/Fe] ratio in \seg\ appears normal with 
respect to all of the comparison samples.
The [Sr/Ba] ratio in \seg\ is $-$0.35~$\pm$~0.27.
For comparison, 
the five highly \rpro\ enhanced field stars 
studied by \citet{sneden09} have a mean [Sr/Ba]
ratio of $-$0.16~$\pm$~0.11.
\citet{aoki08} studied a sample
of eight field stars with [Fe/H]~$< -$2
and high levels of \spro\ enhancement.
In contrast, 
the mean [Sr/Ba] ratio in these stars was found to be
$-$1.17~$\pm$~0.18.
The [Sr/Ba] ratio in \seg\ is in agreement with this
ratio in the \rpro\ enhanced stars
but not with that in the \spro\ enhanced stars.

The [Sr/Ba] ratio in \seg\ also falls within
the normal range for other ultra-faint dwarf galaxies.
This ratio ratio varies by more than 2~dex within
individual ultra-faint dwarf galaxies 
(\umagal; \citealt{frebel10}) and among the ensemble of these systems.
Four stars in \boogal\ where Sr~\textsc{ii} and Ba~\textsc{ii} are detected
indicate [Sr/Ba] is significantly subsolar in that system,
with [Sr/Ba]~$= -$0.9~$\pm$~0.3 \citep{norris10b,ishigaki14}.
Such low ratios are found among halo field stars, but they are rare;
in the \citet{roederer14} field star sample, only 
11~per cent of the red giants have [Sr/Ba]~$< -$0.3.
The growing significance of this discrepancy 
between [Sr/Ba] in field stars and the ultra-faint dwarf galaxies
should be monitored as more data become available.

\section{Discussion}
\label{discussion}

\subsection{Metal Production in Segue 2}
\label{metals}

The [Mg/Fe], [Ca/Fe], and possibly [Ti/Fe] ratios in \seg\
are enhanced relative to the solar ratios
and are not subsolar like those found in the more 
metal-rich stars of classical Local Group dwarf galaxies.
Neither the intermediate odd-$Z$ elements (Na, Al, K)
nor the iron group elements show any significant deviations
from the abundance patterns
commonly found in field stars or dwarf galaxies.
Ratios among the neutron-capture elements and iron ([Sr/Fe], [Ba/Fe])
are subsolar by more than 1~dex, but these deficiencies
are common in the ultra-faint dwarf galaxy populations
and lie at the low ends of the halo star distributions.
This abundance pattern can be attributed to 
enrichment by Type~II supernovae.

\begin{figure}
\begin{center}
\includegraphics[angle=00,width=3.0in]{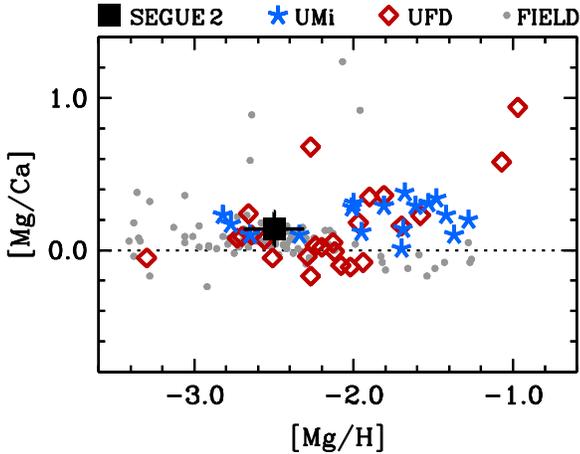}
\end{center}
\caption{
\label{abundplot5}
The [Mg/Ca] ratios as a function of [Mg/H].
Symbols are the same as in Fig.~\ref{abundplot1}.
}
\end{figure}

In Type~II supernovae,
magnesium is produced via hydrostatic carbon and neon burning,
and calcium is produced via oxygen burning during the explosion
(e.g., \citealt{woosley95}).
Fig.~\ref{abundplot5} illustrates that the [Mg/Ca] ratio
in \seg\ is the same as in the majority of stars in the field,
\umigal, and the ultra-faint dwarf galaxies 
with similar [Mg/H] ratios.
Fig.~3 of \citet{feltzing09} and Figures~20 and 21 
of \citet{venn12} demonstrate that the [Mg/Ca]
ratios in several other classical and ultra-faint dwarf galaxies 
match this ratio within a factor of $\approx$~2,
regardless of whether [Mg/Fe] and [Ca/Fe] are 
solar or supersolar.

To place this result in context,
it may be helpful to examine 
stars with non-standard [Mg/Ca] ratios.
In the dwarf galaxy population, these include
one star found in \dragal\ \citep*{fulbright04},
two stars in \hergal\ \citep{koch08}, and
one star in \cargal\ \citep{venn12}.
In these stars, the hydrostatic $\alpha$ elements O and Mg
are enhanced relative to the explosive $\alpha$ element Ca.
These authors
interpreted the enhanced [Mg/Ca] ratios as a consequence of
stochastic sampling of the high end of the
Type~II supernova mass function.
\citet{gilmore13} report a star
in \boogal, \mbox{Boo-119}, with a high [Mg/Ca] ratio.
This star also shows enhanced [C/Fe] \citep{lai11} and [Na/Fe],
suggesting it is a member of the class of 
carbon-enhanced metal-poor stars with
no enhancement of neutron-capture elements.
Such stars have been suggested as 
some of the earliest to have formed from the 
remnants of zero-metallicity Pop~III stars (e.g., \citealt{norris13}).
\citet{feltzing09} reported another star with enhanced [Mg/Ca] 
in \boogal, \mbox{Boo-127}, but that result was not confirmed
by \citeauthor{gilmore13}\ and \citet{ishigaki14}.
The similarity of the [Mg/Ca] ratios in \seg,
the field giants, and most classical dwarfs
suggest that the Type~II supernovae
reflect a relatively well-sampled initial mass function (IMF) in \seggal.
%The fact that [Sr/Fe] and [Ba/Fe] are not as deficient
%as in some stars in \dragal, \hergal\ \citep{koch13}, or \cargal\
%also supports this assertion.

\citet{kirby13} excluded inhomogeneous mixing as the
source of the metallicity spread in \seggal\
on account of the dispersion in the
[Si/Fe] and [Ti/Fe] ratios.
Downward trends in these ratios (and possibly [Mg/Fe])
with increasing [Fe/H]
suggest that star formation in \seggal\ occurred
over a timescale long enough to incorporate the
products of multiple supernovae.
Abundance information for more metal-rich stars in
\seggal\ is limited to the 
[Mg/Fe], [Si/Fe], [Ca/Fe], and [Ti/Fe] ratios
derived by \citeauthor{kirby13} %.
These data may indicate that the additional metals were
manufactured by Type~Ia supernovae, which produce
intrinsically lower [$\alpha$/Fe] ratios.
\citet{mcwilliam13} offer an alternative hypothesis
that better explains the declining [$\alpha$/Fe] ratios
and low ratios of hydrostatic to explosive $\alpha$ elements
in the \sgrgal\ dwarf galaxy.
In this scenario, an extension of that proposed initially
by \citet{tolstoy03} for other classical dwarf galaxies,
a low star formation rate
produces a top-light IMF.
The yields of hydrostatic $\alpha$ elements 
increase with increasing stellar mass, so they will
naturally be deficient in such a scenario.
The extant data are insufficient to
distinguish between these scenarios in \seggal,
but the composition of 
at least one of the most metal-poor stars in \seggal\
is dominated by products of
fairly normal Type~II supernovae.
Either way, the \citeauthor{kirby13}\ abundance data
indicate that \seggal\ experienced self-enrichment,
excluding it from the list of candidates
for the ``one-shot enrichment'' scenario proposed by \citet{frebel12}.

\subsection{The Original Mass of Segue 2}
\label{originalmass}

\seggal\ has a present-day stellar mass of $\approx$~10$^{3}$~\msun\
\citep{kirby13},
assuming $M_*/L_V = 1.2$, 
as is typical for dwarf spheroidal galaxies \citep{woo08}.
The [Sr/Fe] ratio in \seggal\
more closely matches that found in the ultra-faint dwarf galaxies
than in \umigal.
This could be an indication that the original mass of \seggal\ 
was not quite as large as the original mass of \umigal.
%All stars that have been studied in the ultra-faint
%dwarf galaxies show subsolar [Sr/Fe] ratios;
Similarly low [Sr/Fe] ratios are also found, however, in 
a few stars in the more luminous systems
\dragal\ \citep{fulbright04}, 
\hergal\ \citep{koch08}, and
\cargal\ \citep{venn12}.
We consider the low [Sr/Fe] ratio in \seg\
inconclusive regarding the original mass of \seggal.

On one hand, the total mass of metals, $\sim$~0.1~\msun,
in a galaxy as small as \seggal\ is consistent with
the predicted yields of a single zero-metallicity 
supernova (cf.\ \leogal; \citealt{simon10}).
Substantial metal loss from dwarf galaxies seems unavoidable, however,
so multiple supernovae may be necessary to 
account for the small fractions of metals retained,
even in the lowest metallicity stars.
\citet{kirby11c} estimate that this fraction could
be 1~per cent or less for dwarf galaxies
like \umigal.
For a \citet{salpeter55} IMF,
only $\sim$~2~stars with $M >$~8~\msun\ would be expected
for every 10$^{3}$~\msun\ of stars formed.
\citeauthor{tumlinson06}'s (2006) chemical evolution model 
predicts that the average halo star with [Fe/H]~$\sim -$3
and normal abundance ratios
has $\sim$~10 enriching progenitors.
This model might also be applicable to \seg\
because this star has abundance ratios like halo stars
of similar metallicity.
Therefore, \seg\ might also require at least $\sim$~10
enriching progenitors to explain its chemical abundances,
or $\sim$~5 times as many as would be expected from a 
\citeauthor{salpeter55} IMF for a galaxy with
\seggal's current stellar mass.
Consequently, we infer that the stellar mass of \seggal\
was at least $\sim$~5 times greater when it first formed stars
than it is today.

Following a different line of reasoning,
\citet{kirby13} estimated that \seggal\ must have had
$\ga$~150~times its current stellar mass
in order to retain the ejecta of one supernova.
This missing mass could be dark matter or other stars
that are no longer part of \seggal.
%\citeauthor{kirby13}\ present an additional argument
%supporting the hypothesis 
%that \seggal\ may have been substantially
%more massive in the distant past.
\citeauthor{kirby13}\ also note that
\seggal\ does not lie on the (present-day)
luminosity-metallicity relation, which
predicts that a galaxy of \seggal's luminosity
should have a mean metallicity of [Fe/H]~$= -$2.83~$\pm$~0.16
\citep{kirby11b},
a factor of 4~lower than derived by \citet{kirby13}.  % yes, \citet
Therefore
\seggal\ would have shed $\approx$~99.7~per cent
of its original stellar mass if it obeyed the
luminosity-metallicity relation 
at the time it was born.
Presumably this mass loss would have occurred as
\seggal\ fell into the Milky Way halo.
Low surface brightness tidal tails have not yet been
detected around \seggal, and 
unfortunately no proper motions
are available to calculate the orbit of \seggal\
with respect to the Milky Way.

\subsection{The Origin of the Neutron-Capture Elements in Segue 2}
\label{ncaps}

Despite our best efforts, no heavy elements 
except strontium and barium
can be detected in our spectrum of \seg.
In Section~\ref{results}
we presented some typical \rpro\ and \spro\
ratios of [Sr/Ba] for comparison,
and the [Sr/Ba] ratio in \seg\ 
is suggestive of an \rpro\ origin.
Yet the [Sr/Ba] ratio alone is hardly sufficient to 
unambiguously determine what kind of
nucleosynthesis reactions may have produced the
heaviest elements found in \seg.
Strontium may be produced by a myriad of neutron-capture and
charged-particle reactions, and the abundance patterns
resulting from $r$- and \spro\
nucleosynthesis depend on the 
physical conditions at the time of nucleosynthesis.
The barium abundance and 
our upper limit on the europium abundance in \seg\
([Eu/Ba]~$< +$0.70)
cannot exclude the main component of the \rpro\
as exemplified by the abundance pattern in the
metal-poor halo star \cs\ (e.g., \citealt{sneden03};
[Eu/Ba]~$= +$0.65).
Prodigious lead production also did not occur,
and the low [Sr/Fe] and [Ba/Fe] ratios suggest 
that large amounts of \spro\ material were not present
in the gas from which \seg\ formed.
Without additional information it is best to avoid
making any definitive statements regarding the origin 
of the neutron-capture elements.

The unmistakable presence of strontium and barium
in the most metal-poor star known in \seggal\
indicates that at least one neutron-capture nucleosynthesis
mechanism operated prior to the formation of this star.
\citet{roederer13} showed that this is a characteristic of
all other systems that have been studied,
and our observations demonstrate that \seggal\ is no exception.
\citeauthor{tumlinson06}'s (2006) model also 
predicts that $\sim$~90~per cent of the supernova progenitors of
the average halo star with [Fe/H]~$\sim -$3
are zero-metallicity progenitors.  
While this prediction remains unverified,
the presence of neutron-capture elements in \seggal\ and 
all other systems 
hints that neutron-capture reactions may have occurred in
at least some zero-metallicity stars
(cf.\ \citealt{roederer14b}).

\section{Summary}
\label{summary}

We have performed a detailed abundance analysis of
the brightest red giant star in the \seggal\ galaxy
using high quality optical spectroscopy obtained
with the MIKE spectrograph.
The fundamental new insight from our analysis is that the
composition of this star is not substantially
different from the majority of stars in
other ultra-faint dwarf galaxies or the most metal-poor stars
in the classical dwarf galaxies like \umigal.
This suggests that multiple
Type~II supernovae were responsible for producing the 
metals observed in \seg.
For a standard \citet{salpeter55} IMF,
this implies that the stellar mass in \seggal\ 
was at least $\sim$~5~times greater than at the present.
Our results echo those of \citet{frebel10} and \citet{simon10}
that the light element abundance patterns in many of 
the ultra-faint dwarf galaxies
generally match those found in halo stars.
The exceptions appear to be the
neutron-capture elements, which are persistently 
deficient in the ultra-faint dwarf galaxies.

\section*{Acknowledgments}

I.U.R.\ thanks J.\ Sobeck and C.\ Sneden
for their expert assistance with \textsc{moog}.
We thank the referee for providing helpful suggestions on the manuscript.
This research has made use of NASA's 
Astrophysics Data System Bibliographic Services, 
the arXiv preprint server operated by Cornell University, 
the SIMBAD and VizieR databases hosted by the
Strasbourg Astronomical Data Center, and 
the Atomic Spectra Database \citep{kramida13} hosted by
the National Institute of Standards and Technology. 
\textsc{iraf} is distributed by the National Optical Astronomy Observatories,
which are operated by the Association of Universities for Research
in Astronomy, Inc., under cooperative agreement with the National
Science Foundation.
E.N.K. acknowledges support from the Southern California Center for
Galaxy Evolution, a multicampus research program funded by the
University of California Office of Research, and partial support from
NSF grant AST-1009973.

% {\it Facilities:} 
%\facility{Magellan:Clay (MIKE)}, 

\label{lastpage}

\end{document}